\documentstyle[defns,amssymbols,12pt]{article}
\begin{document}
\title{\bf Temperature Correlations of Quantum Spins}
\author{\bf A.R. Its$^{1}$, A.G. Izergin$^{2,*}$, V.E. Korepin$^3$, N.A.
Slavnov$^{4}$}
\date{ \ \ }
\maketitle

\vfill
\centerline{\bf INS \#211}

\centerline{\bf ITP-SB-92-47}

\centerline{\bf ENSLAPP 394}

\vfill

\begin{abstract}
We consider isotropic XY model in the transverse magnetic field in one
dimension.  One
can alternatively call this model Heisenberg XXO antiferromagnet.  We solve
 the
problem of evaluation of asymptotics of temperature correlations and
explain the
physical meaning of our result.  To do this
we represent quantum correlation function as a tau function of a completely
integrable
differential equation.  This is the well-known Ablowitz-Ladik lattice
nonlinear \s
differential equation.
\end{abstract}

\vfill

\noindent $^1$Department of Mathematics and Computer Science
and the Institute for Nonlinear Studies
Clarkson University
Potsdam, New York  13699-5815, U.S.A.

\noindent $^2$Laboratorie de Physique Th\'eorique ENSLAPP
Ecole Normale, Superieure de Lyon, France

\noindent $^3$Institute for Theoretical Physics
State University of New York
Stony Brook, NY  11794-3840, U.S.A.
e-mail korepin@max.physics.sunysb.edu
\noindent $^{*}$On leave of absence from St. Petersburg Branch (POMI) of
Steklov Mathematical Institute of Russian
Academy of Science
Fontanka 27, 1191011 St. Petersburg, Russia

\noindent $^{4}$V.A. Steklov
Mathematical Institute of Russian Academy of Science
Moscow, Russia

\vfill
\vfill

\pagebreak

\baselineskip 20pt

XY model was introduced and studied by E. Lieb, T. Schultz and D. Mattis
[1].  It
describes interaction of spins 1/2 situated on 1-dimensional periodic
lattice.  The
Hamiltonian of the model is

$$ H =- \sum_n\left[\sigma^x_n \sigma^x_{n+1} + \sigma^y_n\sigma^y_{n+1} +
h\sigma^z_n\right]. \eqno (1)$$
Here $\sigma$ are Pauli matrices, $h$ is transverse magnetic field, $n$
enumerate the
cites of the lattice.  At zero temperature the problem of evaluation of
asymptotics of
correlation functions was solved in [2,3].  Here we consider temperature
correlation
function

$$ g(n,t) = \frac{Tr
\{e^{-\frac{H}{T}}\sigma^+_{n_2}(t_2)\sigma^-_{n_1}(t_1)\}}{Tr
e^{-\frac{H}{T}}}, \ \ \ \
n = n_2-n_1, \ \ \ \ t=t_2-t_1 \eqno (2) $$
for the infinite lattice.

We consider finite temperature $0 < T < \infty$ and moderate magnetic field
$0 \leq h <
2$.

We evaluated asymptotics in cases where both space and time separation go
to infinity $n
\rightarrow \infty$, $t\rightarrow \infty$, in some direction $\varphi$

$$ \frac{n}{4t} = \cot \varphi, \ \ \ \ 0 \leq \varphi \leq \frac{\pi}{2}.
\eqno (3)$$
In accordance with our calculations, correlation function $g(n,t)$ decays
exponentially in any direction, but the rate of
decay depends on the direction.  In the space like direction $0 \leq
\varphi <
\frac{\pi}{4}$ asymptotics is

$$ g(n,t) \rightarrow C \exp \left\{ \frac{n}{2\pi}\int^\pi_{-\pi}dp\ln
\left|\tanh
(\frac{h-2\cos p}{T})\right|\right\}. \eqno (4)$$
In the time like direction $\frac{\pi}{4} < \varphi \leq \frac{\pi}{2}$
asymptotics is
different.

$$ g(n,t) \rightarrow C t^{(2\nu^2_++2\nu^2_-)}\exp \left\{
\frac{1}{2\pi} \int^\pi_{-\pi}dp
\left|n-4t\sin p\right|\ln\left|\tanh(\frac{h-2\cos p}{T})\right|\right\}.
\eqno (5)$$
The values $\nu_\pm$, which define pre-exponent, are

$$ \nu_+ = \frac{1}{2\pi}\ln\left|\tanh \left( \frac{h-2\cos
p_0}{T}\right)\right|$$

$$ \nu_- = \frac{1}{2\pi}\ln\left|\tanh \left(
\frac{h+2\cos\rho_0}{T}\right)\right|\eqno (6)$$
where $\frac{n}{4t} = \sin p_0$.  Formula (5) is valid in the whole time
like cone,
with exception of one direction $h=2\cos p_0$.  Higher asymptotic
corrections will
modify formulae by factor $(1+c(t,x))$  ($c$ decays exponentially in the
space-like
region and as $t^{-1/2}$ in the time-like region).  Also, it should be
mentioned that the
constant factor $C$ in (4) does not depend on the direction $\varphi$, but
it does depend on
$\varphi$ in (5).

We want to emphasize that for pure time direction $\varphi = \pi/2$ leading
factor in
asymptotics (exponent in (5)) was first obtained in [10].

To
derive these formulae we went through few steps.

The first step: The explicit expression for eigenfunctions of the
Hamiltonian (1) (see [1])
was used to represent the correlation function as a determinant of an
integral
operator (of Fredholm type) [8].  In order to explain we need to introduce
notations.
Let us consider integral operator $\hat V$.  Its kernel is equal to

$$ V(\lambda\mu) = \frac{e_+(\lambda)e_-(\mu) -
e_-(\lambda)e_+(\mu)}{\pi(\lambda -
\mu)}. \eqno (8)$$
Here $\lambda$ and $\mu$ are complex variables, which go along the  circle
$\left|\lambda\right| = \left|\mu\right| = 1$ in a positive direction.
Functions $e_\pm$ are

$$ e_-(\lambda) = \lambda^{-n/2} \cdot e^{-it(\lambda+1/\lambda)}
\sqrt{v(\lambda)} \eqno
(9)$$
where

$$ v(\lambda) = \left\{ 1 + \exp \left[ \frac{2h-2(\lambda +
\frac{1}{\lambda})}{T}\right]\right\}^{-1} \eqno (10)$$
and

$$ e_+(\lambda) = e_-(\lambda)E(x,t,\lambda). \eqno (11)$$
Here $E$ is defined as an integral

$$ E(n,t,\lambda) = \frac{1}{\pi}v.p.\int\exp\{ 2it(\mu + \frac{1}{\mu})\}
\cdot
\frac{\mu^nd\mu}{\mu-\lambda}. \eqno (12)$$
It is convenient to define functions $f_\pm(\lambda)$ as solutions of the
following
integral equations:

$$ (I + \hat V)f_k = e_k. \eqno (13)$$
Here $I$ is an identical operator and $k =\pm$.  Next we define potentials
$B_{kj} (k,j
= \pm)$:

$$ B_{kj}(n,t) = \frac{1}{2\pi i} \int f_k(\lambda)
e_j(\lambda)\frac{d\lambda}{\lambda}.
\eqno (14)$$
They depend on space and time variables $n,t$.  These we shall use to
define new
potentials $b_{kj}$:

$$ \begin{array}{ll} b_{--}(n,t) = B_{--}(n,t) \\ \\

 b_{++}(n,t) = B_{++}(n,t) - 2iG(n,t)B_{+-}(n,t) - G(n,t). \end{array}
\eqno (15)$$

\noindent Here we used function

$$ G(n,t) = \frac{1}{2\pi i}\int \lambda^{n-1}\exp \{ 2it(\lambda +
\frac{1}{\lambda})\}
d\lambda. \eqno (16)$$
Now all the notation are ready to write determinant formula
for the correlation function $g(n,t)$ (see (4)):

$$ g(n,t) = e^{-2iht}b_{++}(n,t)\exp \{ \sigma(n,t)\}. \eqno (17)$$
Here $e^\sigma$ is a determinant of integral operator

$$ \exp\{ \sigma(n,t)\} = \det(1+\hat V). \eqno (18)$$

Second step: Formulae (8)-(15) can be used to show that potentials $b_{++}$
and $b_{--}$
satisfy the system of nonlinear differential equations.

$$ \frac{i}{2} \frac{\partial}{\partial t} b_{--}(n,t) = (1 +
4b_{--}(n,t)b_{++}(n,t))(b_{--}(n+1,t) + b_{--}(n-1,t)) $$

$$ -\frac{i}{2} \frac{\partial}{\partial t} b_{++}(n,t) =
(1+4b_{--}(n,t)b_{++}(n,t))(b_{++}(n+1,t) + b_{++}(n-1,t)). \eqno (19)$$
Derivation of these equations is similar to [4,5,7].  Equations (19) are
completely
integrable differential equations.  They were first discovered by Ablowitz
and Ladik [9]
as an integrable discretization of the nonlinear \s equation.  Logarithmic
derivatives
of $\sigma(x,t)$ (see (18)) can be expressed in terms of solutions of the
system (19).

$$ \begin{array}{ll}
\frac{\partial^2\sigma(n,t)}{16\partial t^2} & = 2b_{--}(n,t)b_{++}(n,t) -
b_{++}(n-1,t)b_{--}(n+1,t) - \\ \\

& -b_{--}(n-1,t)b_{++}(n+1,t) -
4b_{++}(n,t)b_{--}(n,t)[b_{++}(n-1,t)b_{--}(n+1,t) +\\
\\

& b_{--}(n-1,t)b_{++}(n+1,t)] \end{array} \eqno (20)$$

$$ \sigma(n+1,t) + \sigma(n-1,t) -2\sigma(n,t) =
\ln[1+4b_{--}(n,t)b_{++}(n,t)] \eqno
(21)$$

$$ \frac{\partial}{\partial t}[\sigma(n+1,t) - \sigma(n,t)] =
8i[b_{++}(n+1,t)b_{--}(n,t) - b_{++}(n,t)b_{--}(n+1,t)]. \eqno (22)$$
This shows that quantum correlation function $g$ (2) can be expressed in
terms of
solution of the system (19).

The meaning of all these formulae is that correlation function of $XY$
model is the
$\tau$ function of Ablowitz-Ladik's differential-difference equation.  This
is similar
to quantum nonlinear Shrodinger equation.  In the papers [5,6,7] relation
of $\tau$ function and
quantum correlation functions is explained in more detail.

Third step: In order to evaluate asymptotics one should solve
Ablowitz-Ladik's
differential equation.  Initial data can be extracted from integral
representations
(8)-(18).  We use Riemann-Hilbert problem in order to evaluate asymptotics
of solution
of equation (19).  It is quite similar to the nonlinear Shrodinger  case
[6,7].

Let us emphasize that it is quite remarkable that for the $XY$ model (like
for Bose gas, [4-7] which is defined by quantum nonlinear Shrodinger
equation) quantum
correlation function is the $\tau$ function of classical completely
integrable
differential equation.

In the end let us explain the physical meaning of our asymptotic formula
(4).  We start
from the expression for free energy [1]:

$$ f(h)=-h-\frac{T}{2\pi}\int^\pi_{-\pi}dp \ln\left(1 + \exp[ \frac{4\cos
p-2h}{T}]\right).  \eqno (23)$$
We emphasize the dependence on the magnetic field $h$.  The definition of
$f(h)$ is
standard:

$$ Tre^{-\frac{H}{T}} = \exp\{ -\frac{L}{T}f(h)\}. \eqno (24)$$
Here $L$ is the length of the box.  Let us use Jordan-Wigner transformation
to transform
correlator (2) (in the equal time case):

$$ \sigma^+_{n_2}(0)\sigma^-_{n_1}(0) = \psi_{n_2}\exp \left\{ i\pi
\sum^{n_2-1}_{k=n_1+1}\psi^+_k\psi_k\right\}\psi^+_{n_1}, \ \ \ \
\psi^+_k\psi_k=\frac{1}{2}(1-
\sigma^z_k). \eqno (25)$$
Here $\psi_k$ is canonical Fermi field.  We can say now that numerator in
(2) differs
from denominator by replacement of magnetic field $h\rightarrow h-i\pi T/2$
on the space
interval $[n_1+1,n_2-1]$.  This leads us to the following asymptotical
expression for
correlator $g(n,0)$

$$ g(n,0) \rightarrow \exp\left\{ Re \frac{n}{T} \left[ f(h)-f(h-
\frac{i\pi T}{2})\right]\right\}. \eqno (26) $$
The reason why we wrote $Re$ is that $i\pi$ in (25) can be replaced by
$-i\pi$.
It is remarkable that (26) coincides with the correct answer (4).  It is
also worth
mentioning that to go to exponent in (5) one should replace differential
$d(np)$ by
expression $|d(np-t\varepsilon(p)|$ where $\varepsilon(p) =-4\cos p+2h$ is
the energy of
the quasiparticle of the model.

\vskip .2in
\noindent{\Large Acknowledgements}

This work was partially supported by NSF Grant No. PHY-9107261 and by NATO
S.P. CRG
901098.  One of the authors (A.G.I.) is grateful to the Laboratoire de
Physique
Theorique in Ecole Normale Superrieure de Lyon (France) for their warm
hospitality.

\end{enumerate}


\begin{thebibliography}{5}
\bibitem{1} E. Lieb, T. Schultz and D. Mattis, Ann. Phys. (N.Y.) {\bf 16},
406 (1961).

\bibitem{2} B.M. McCoy, J.H.H. Perk, R.E. Shrock, Nucl. Phys. B {\bf 220},
269 (1983);
Nucl. Phys. B {\bf 220}, 35 (1983).

\bibitem{3} H.G. Vaidya, C.A. Tracy, Phys. Lett. A {\bf 68}, 378 (1978).

\bibitem{4} A.R. Its, A.G. Izergin, V.E. Korepin, N.A. Slavnov, J. Mod.
Phys. B. {\bf
4}, 1003 (1990).

\bibitem{5} A.R. Its, A.G. Izergin, V.E. Korepin, N.A. Slavnov, in the book
``Important
Developments in Soliton Theory, 1980-1990", eds. A.S. Fokas, V.E. Zakharov,
Springer-Verlag, 1992.

\bibitem{6} A.R. Its, A.G. Izergin, V.E. Korepin, G.G. Varzugin, Physica D,
{\bf 54},
351 (1992).

\bibitem{7} V.E. Korepin, A.G. Izergin, N.M. Bogolinbov, ``Quantum Inverse
Scattering
Method, Correlation Functions and Algebraic Bethe Ansatz" Cambridge
University Press,
(1992).

\bibitem{8} F. Colomo, A.G. Izergin, V.E. Korepin, V. Tognetti,
``Determinant
Representation for Correlation Functions in the XXO Heisenberg Chain"
preprint
ITP-SB-92-12, accepted in Phys. Lett.

\bibitem{9} M.J. Ablowitz, J.F. Ladik, Stud. Appl. Math. {\bf 55}, 213
(1976).

\bibitem{10} P. Deift, X. Zhou, preprint, Courant Mathematical Institute
(1992).
\end{thebibliography}
\end{document}